\documentclass[aps,showpacs,prl,twocolumn,superscriptaddress,letter]{revtex4}
\usepackage{graphicx}

\begin{document}
\newcommand{\pH}{p-H$_2$}
\newcommand{\oD}{o-D$_2$}

\bibliographystyle{apsrev}

\preprint{}

\title[]{Study of distortion effects and clustering of isotopic impurities
in solid molecular para-hydrogen by Shadow Wave Functions}

\author{Francesco Operetto}
\affiliation{Dipartimento di Fisica and INFN, Universit\`a di Trento, via Sommarive, 14
I-38050 Povo, Trento, Italy
}

\author{Francesco Pederiva}
\affiliation{Dipartimento di Fisica and INFN, Universit\`a di Trento, 
via Sommarive, 14 I-38050 Povo, Trento, Italy
}
\affiliation{INFM/CNR DEMOCRITOS National Simulation Center, Trieste, Italy}

\date{\today}

\begin{abstract}
We employed a fully optimized Shadow Wave Function (SWF) in combination 
with Variational Monte Carlo techniques to investigate the properties
of HD molecules and molecular ortho-deuterium (\oD) in bulk solid 
para-hydrogen (\pH).
Calculations were performed for different concentrations of impurities 
ranging from about 1\% to 25\% at the equilibrium density for
the para-hydrogen crystal.
By computing the excess energy both for clustered and isolated 
impurities we tried to determine a limit for the solubility 
of HD and \oD\ in \pH. 
\end{abstract}

\maketitle

%\section{Introduction}

Isotopic impurities in quantum crystals are often used as a marker for 
testing 
the occurrence of quantum mechanical particle--particle 
exchange by means of the NMR experiments\cite{Ral92,Kis94}. In particular, 
in the case of $J=0$ impurities, like HD or (\oD) molecules, 
embedded in a $J=0$ para-hydrogen (\pH) matrix with  small concentrations
of ortho-hydrogen (less than 3\%), the observed longer time relaxation
  can only be attributed 
to particle--particle tunneling, because the so--called ``resonant conversion'' is not allowed
(see Ref. \onlinecite{Kis94} and its inner references). 
Concentrations of HD molecules in these   experiments are  about 1-2\%\cite{Ral92}. 
It is safe to assume that the interaction among different hydrogen isotopes are  
very similar. Only the mass difference should 
distinguish the behavior of \pH\ and HD. This is the main reason for
relating  the motion of HD impurities
to the motion of the molecules in the matrix.
However, in a quantum mechanical solid, the mass difference among
the constituents can act in quite subtle way. In fact, heavier particles
tend to be more localized,
with a net gain in energy, while 
their nearest neighbors might find convenient to
lower their kinetic energy by relaxing towards the impurity, in 
a very delicate balance between the gain in kinetic energy and the loss 
in potential energy.
%As a consequence, their nearest neighbors might find convenient to
%lower their kinetic energy by relaxing towards the impurity, in 
%a very delicate balance with a possible loss in potential energy.
This is the case of HD and \oD impurities in H$_2$. The presence of 
impurities of lower mass (like $^3$He in $^4$He) should instead have a less
dramatic effect, due to  steric hindrance which prevents the impurity from
relaxing too much against the rest of the crystal. 
Change of force constants and lattice distortion in the proximity of a isotopic impurity were studied
in relation to their influence on the thermal conductivity of solid \pH\ at low temperatures\cite{Kor99}.
More recently, Raman spectroscopy experiments\cite{Koz03} clearly showed
an effect on both rotational and vibrational spectra which can
be attributed to local relaxation of the lattice around the impurity.
The balance among kinetic and potential energy might also
lead to clustering phenomena when the concentration of impurities
exceeds a certain limit. Finally, HD impurities might play a role in 
the determining the results on measurements of non classical torsional
inertia recently performed to investigate the existence of a supersolid
phase in \pH\cite{Cla06}.

In this  letter we report the results of Variational Monte Carlo (VMC)
simulations performed with a modified Shadow Wave Function\cite{Vit88,Rea88}
(SWF) to study the properties of HD and \oD\ molecules in a pure \pH\ matrix.
The main property of SWF is the capability of describing both the 
crystalline and the disordered phase of a quantum system with the same 
functional form, with is always translationally invariant. Unlike in the standard 
Jastrow--Nosanow trial wave function, using SWF no {\em a priori} equilibrium are 
imposed to describe the crystalline phase. This property
allows SWF to provide a realistic description of a variety of inhomogeneous
system, like liquid--solid phase coexistence in $^4$He\cite{Ped94}, as well 
as finite system as, for instance, $^4$He clusters with and without impurities\cite{Gal01}.
Recently a SWF has been employed also for \pH, in order to describe the homogeneous 
liquid and solid phase\cite{Ope04}, as well as to investigate energetic and 
structural properties of defective crystals\cite{Ope06_vac}.

%\section{Methods}
%\subsection{Hamiltonian}
We model the system of $N_H$ molecules of \pH\ and $N_I$ isotopic impurities as
a set of $N=N_H+N_I$ point particles described by the following Hamiltonian:
\begin{equation}\label{eq:ham}
\mathcal{H}=-\frac{\hbar^2}{2m_H}\sum_{i=1}^{N_H}\mathbf{\nabla}^2_i
-\frac{\hbar^2}{2m_I}\sum_{i=1}^{N_I}\mathbf{\nabla}^2_i
+\sum_{i<j}v(r_{ij}),
\end{equation}
where $v(r)$ is the well--known Silvera--Goldman (SG) potential\cite{Sil78}.
This model interaction includes a term $\propto 1/r^9$ which effectively accounts
for triple dipole interactions in the system.
This approximation is supported by the fact that
both the \pH\ molecule and the isotopic impurities considered in this paper, HD and \oD\ are
in the rotational ground state ($J=0$), and therefore the average interaction 
must be spherically symmetric. Higher terms in a multipolar expansion, due to the
non--spherical nature of the molecules,  may be safely considered as  second order.
The limit of validity of the SG potential and the influence of explicit many--body 
terms in the potential on the equation of state have been recently tested 
both for solid \pH\cite{Ope06_eos_h2} and \oD\cite{Ope06_eos_d2}.
In the latter case, 
the SG potential has been found to give a rather good description of the equation of state, while
in the case of solid \pH\ the same interaction leads to a slightly more inaccurate result on the energies 
due to stronger many--body effects. However, the explicit angular dependence of the
potential  seems not to have sizeable effects on the 
structural properties, such as the pair distribution function. 
%\subsection{Shadow Wave Function}
The many-body quantum mechanical problem is solved at temperature T=0
by means of a trial solution for the ground--state of the Hamiltonian in Eq. \ref{eq:ham}
in the form of a modified Shadow Wave Function:
\begin{equation}\label{eq:swf}
\Psi_T(R)=\phi_r(R)\int\!K(R,S)\phi_s(S)\textrm{d}S,
\end{equation}
where $R=\{{\bf r}_1\dots{\bf r}_N\}$ are the coordinates of the molecules 
and $S=\{{\bf s}_1\dots{\bf s}_N\}$ are $3N$ auxiliary degrees of freedom
named {\em shadows}.
Both functions \(\phi_r\)\ 
and \(\phi_s\)\ have been expressed as Jastrow products of pair
functions: 
\begin{eqnarray}
\phi_\alpha(X)&=&
\exp\left[\frac{1}{2}\sum^{N_H}_{i<j}u_{HH}(x_{ij})\right]
\exp\left[\frac{1}{2}\sum^{N_I}_{i<j}u_{II}(x_{ij})\right]\times \nonumber\\ \\ 
&&\times \exp\left[\frac{1}{2}\sum^{N_H}_{i=1}\sum^{N_I}_{j=1}u_{HI}(x_{ij})\right],\nonumber
\end{eqnarray}
while the kernel $K(R,S)$ is written as a product of
Gaussians connecting the real and auxiliary degrees of freedom:
\begin{equation}
K(R,S)=\prod_{i=1}^{N_H}\exp\left[-C_H(\mathbf{r}_i-\mathbf{s}_i)^2\right]
\prod_{i=1}^{N_D}\exp\left[-C_I(\mathbf{r}_i-\mathbf{s}_i)^2\right].
\end{equation}
The pair pseudopotentials between molecules have been expanded in term of suitable 
basis functions:
\begin{equation}
u_{xy}(r)=\left(\frac{b_{xy}}{r}\right)^5+\sum_ma_{xy}^m\chi_m(r),
\end{equation}
where both $x$ and $y$ indicate generically one of the two label ''$H$'' or ''$I$''. 
The basis functions $\chi_n$ (about 40) are the same used in Ref. \onlinecite{Ope06_eos_h2}
and \onlinecite{Ope06_eos_d2} for DMC calculations in solid \pH\ and \oD.
\begin{equation}
\chi_m(r)=\left\{\begin{array}{ll}
\Big\{1-\cos\Big[\frac{2\pi m}{L/2-r_c}(r-L/2)\Big]\Big\}r^{-5} & 
\textrm{\ } r>r_c\\
0 & \textrm{\ } r\le r_c \end{array}, \right .
\end{equation}
where $L$ is the side of the simulation box.
The cutoff radius $r_c$ allows to remove  divergences for $r\to0$ in order to 
avoid numerical instabilities in the optimization procedure.
If such cutoff is small enough, it does not influence the value of the energy.
In our calculations it has been taken equal to 1\AA. 
The pseudopotential between auxiliary degrees of freedom is indeed 
just the rescaled Silvera--Goldman interaction:
\begin{equation}
\widetilde{u}_{xx}(s)=\delta_{xx}v(\alpha_{xx}s).
\end{equation}
The parameters $\{b_{xy}\}$, $\{a_{xy}^m\}$, $\{C_{x}\}$, $\{\alpha_{xy}\}$,  
$\{\delta_{xy}\}$ have been optimized following the reweighting scheme, which 
alternate between VMC simulations and minimization of a combination of the variance 
on the expectation value of the Hamiltonian, and of the expectation value itself 
estimated on a subset of the sampled configurations.
%\subsection{Simulations}
We performed simulations for a face centered cubic (fcc) lattice at densities 
$\rho=0.02609\AA^{-3}$, very close to the bulk equilibrium density of pure \pH\cite{Ope06_eos_h2}.
The simulation cell was set to accommodate $3\!\times\!3\!\times\!3$ elementary cubic cells 
for fcc, with a total of $N$=108 lattice sites.
All the intermolecular interactions are truncated at the edge of a sphere of radius equal to 
$L/2$, where $L$ is the length of the side of the cell. 
The contribution from the potential energy outside the sphere is estimated 
by integrating the potential in the $(L/2,+\infty)$ interval, assuming
therefore all the pair correlation functions of molecules and impurities to be constant 
beyond $L/2$. 
Calculations were performed for different number of impurities, from 1 to 27, corresponding
to concentrations ranging from about 1\% to 25\%. 
For each value of concentration, we repeated the optimization of the trial wave function
both with all the impurities initially localized around next--neighbours site, as close
to each other as possible, and with far impurities, {\em i. e.}, separated by at least 
a shell of \pH.
%\section{Results}
The excess energy per impurity $\epsilon_{imp}$ is defined as follow:
\begin{equation}
\epsilon_{imp}=\frac{1}{N_I}\left[E(N_H,N_I)-E(N,0)\right],
\end{equation}
where $E(N_H,N_I)$\ is the total energy of the system of $N_H$ molecules of \pH\ and
$N_I=N-N_H$ impurities.
\begin{figure}
\includegraphics[scale=0.30]{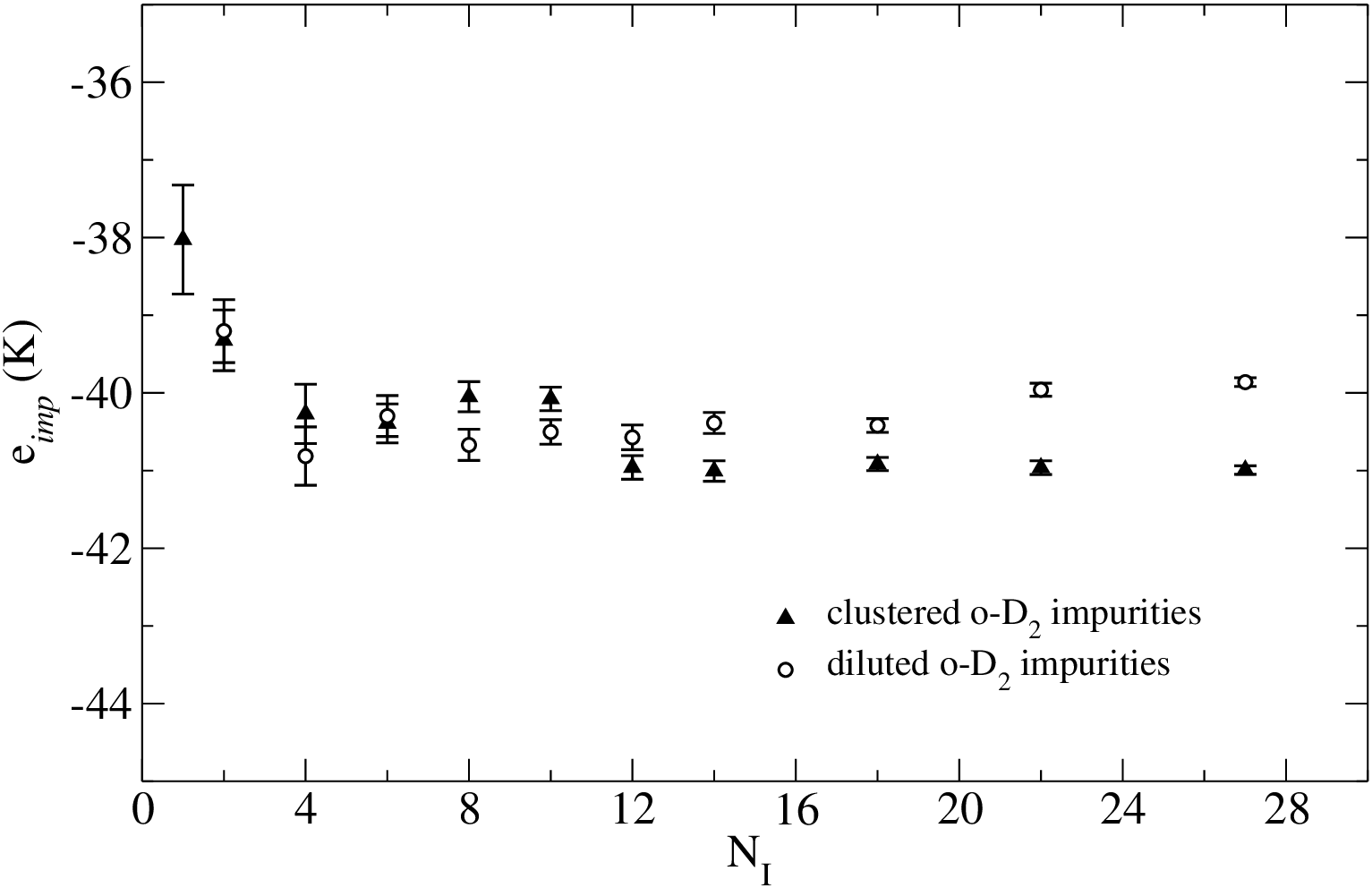}
\caption{Excess energy per impurity (in K) in the \pH\ crystal as function of the number of \oD\ impurities. 
Filled triangles: close impurities, empty circles: far impurities.}
\label{fig:en_d2}
 \end{figure}
\begin{figure}
\includegraphics[scale=0.30]{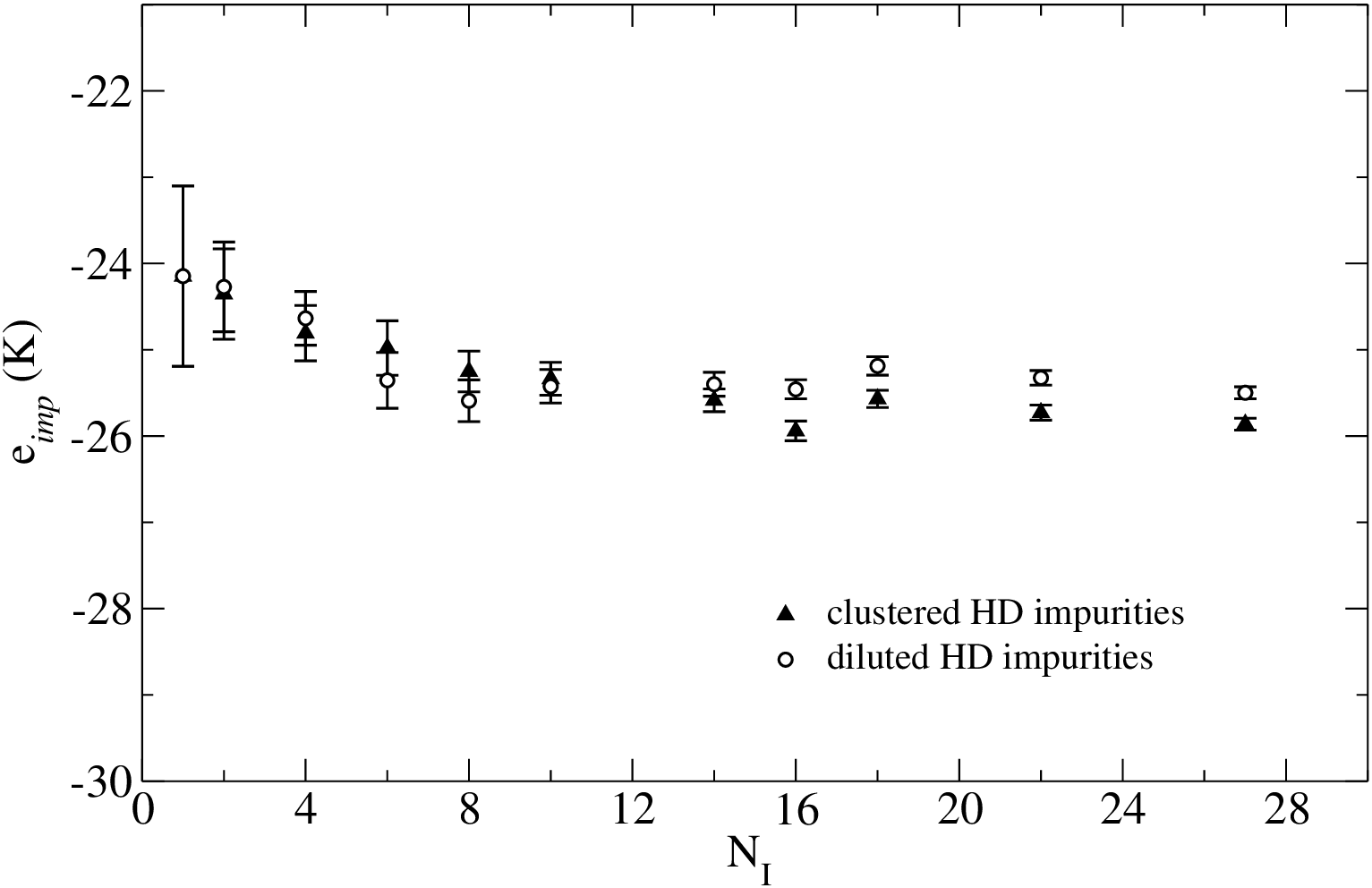}
\caption{Excess energy per impurity (in K) in the \pH\ crystal as function of the number of HD impurities. 
Filled triangles: close impurities, empty circles: far impurities.}
\label{fig:en_hd}
 \end{figure}
It can be noticed that the excess energy per impurity
remains almost constant in the range of concentrations considered. 
A rough estimate of the excess energy 
of a single isotopic impurity in the \pH\ crystal might be obtained considering as a 
perturbation term in the Hamiltonian the difference 
in kinetic energy $\Delta E_{kin}$ due to the mass difference of the two 
isotopes. 
\begin{equation}
e_{imp}\approx\left(\frac{m_H}{m_D}-1\right)\Delta E_{kin}
\end{equation}
The excess binding energy estimated in this way is about 35K for \oD\,  and about 24K for HD. 
Both  values are not far from the values of about 40K and 25K respectively,
obtained from the VMC simulations. 
This fact suggests that the contribution of the 
mass difference to the excess energy of the impurity is dominant,
in particular for HD impurities. We point out that in SWF simulations
both relaxation and possible quantum diffusion effects are correctly taken
into account.
As shown in Figures \ref{fig:en_d2} and \ref{fig:en_hd},
the excess energy computed with clustered or diluted impurities, 
is essentially coincident within errorbars for low concentrations. 
For concentrations up to 10\% we are therefore lead
to consider the \oD\ as not subject to clustering in the
\pH\ crystal matrix. 
For higher concentrations, however, the excess energy
per impurity for clustered molecules becomes lower 
than the energy for the diluted ones well outside errorbars.
In the case of HD impurities, this threshold is higher,
as it might be expected because of the lighter mass, and the VMC 
results are compatible with a limiting concentration for clustering of about 15\%.
In order to check if the estimate of the excess energy
and of the clustering limits are affected either by finite--size effects or
by the choice of the lattice fitted by the simulation box,
 we performed some simulations in a larger  fcc lattice using 
$N=256$ molecules arranged on $4\!\times\!4\!\times\!4$ elementary cubic cells
and in the hcp lattice using $N=180$ 
molecules filling $5\!\times\!3\!\times\!3$ elementary cells.
 In this case the simulation box is not cubic, but
the ratio of the three sides is as close as possible to 1.
The results for concentration of 25\% of 
\oD\ molecules obtained with different lattices and different values of $N$ %are reported in Tab. \ref{tab:n}.
show that the values of the excess energy are largely
independent from from the total number of particles, indicating the absence of
substantial size--finite effects.
%\begin{table}
%\begin{tabular}{c|c|c}
%\toprule
%$N$ &   $e_{imp}$ {\scriptsize close} & $e_{imp}$ {\scriptsize far} \\
%\hline
%108 (fcc) &  -40.95(8)    &    -39.82(8) \\
%180 (hcp) &  -40.85(8)    &    -39.60(8) \\
%256 (fcc) &  -40.71(8)    &    -39.81(8) \\
%\botrule
%\end{tabular}
%\caption{Excess energies per impurity for a concentrations of 25\%
%of clustered (''close'') or diluted (''far'') \oD\ molecules. Calculation were obtained with 
%different lattices and numbers of total particles. Energies are given in Kelvin.
%\label{tab:n}
%}
%\end{table}

\begin{figure}
\includegraphics[scale=0.30]{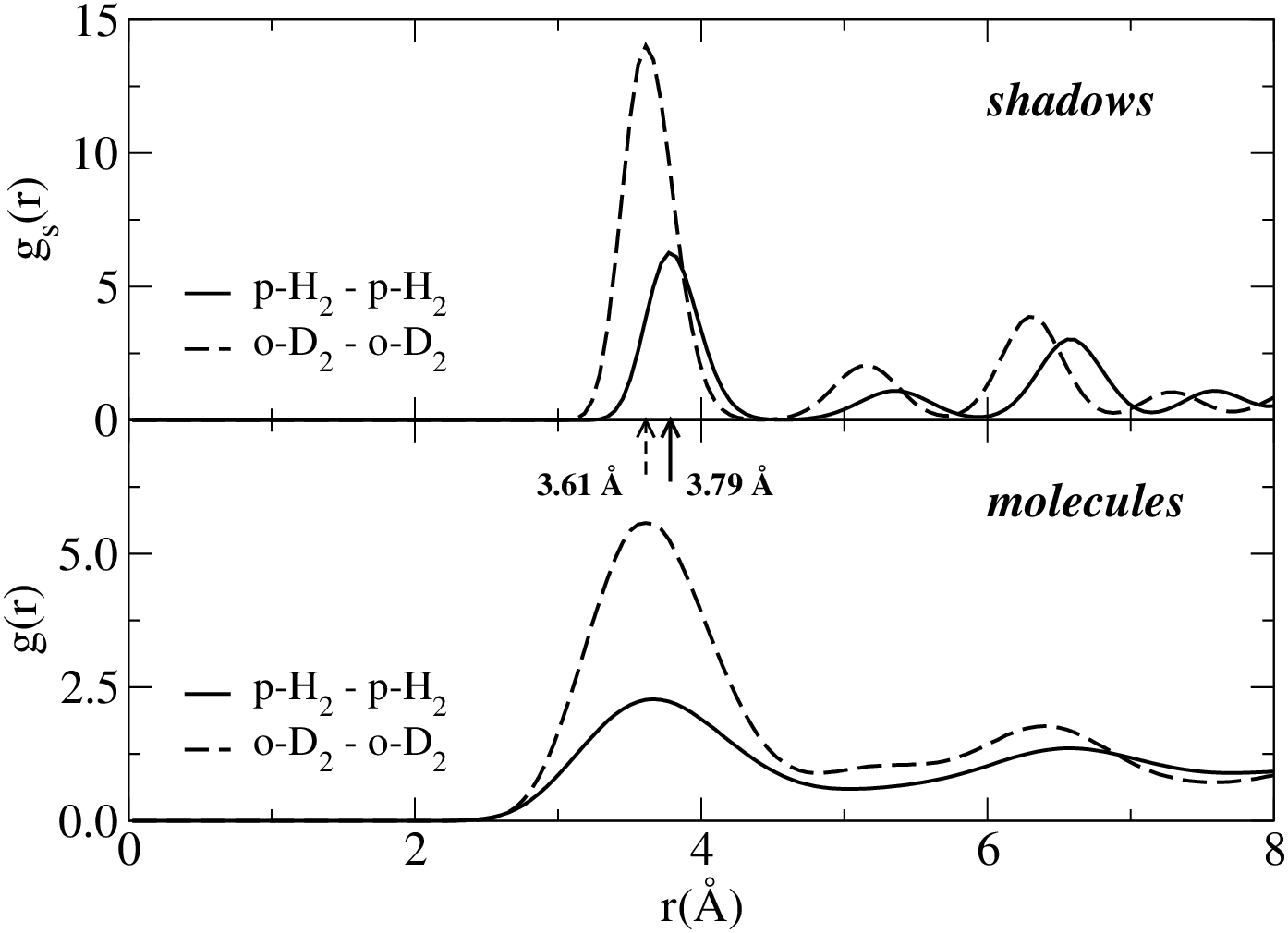}
\caption{Pair correlations functions $g_{\alpha\beta}(r)$ and $g_{s\alpha\beta}(r)$ for $\alpha=\beta=$\pH\ and
$\alpha=\beta$=\oD\  for a 25\% concentration
of clustered impurities in a fcc lattice. Solid and dashed arrows indicate the positions of the next--neighbours peaks
for \pH\ and \oD\ molecules, respectively.} 
\label{fig:gr_d2}
 \end{figure}

\begin{figure}
\includegraphics[scale=0.30]{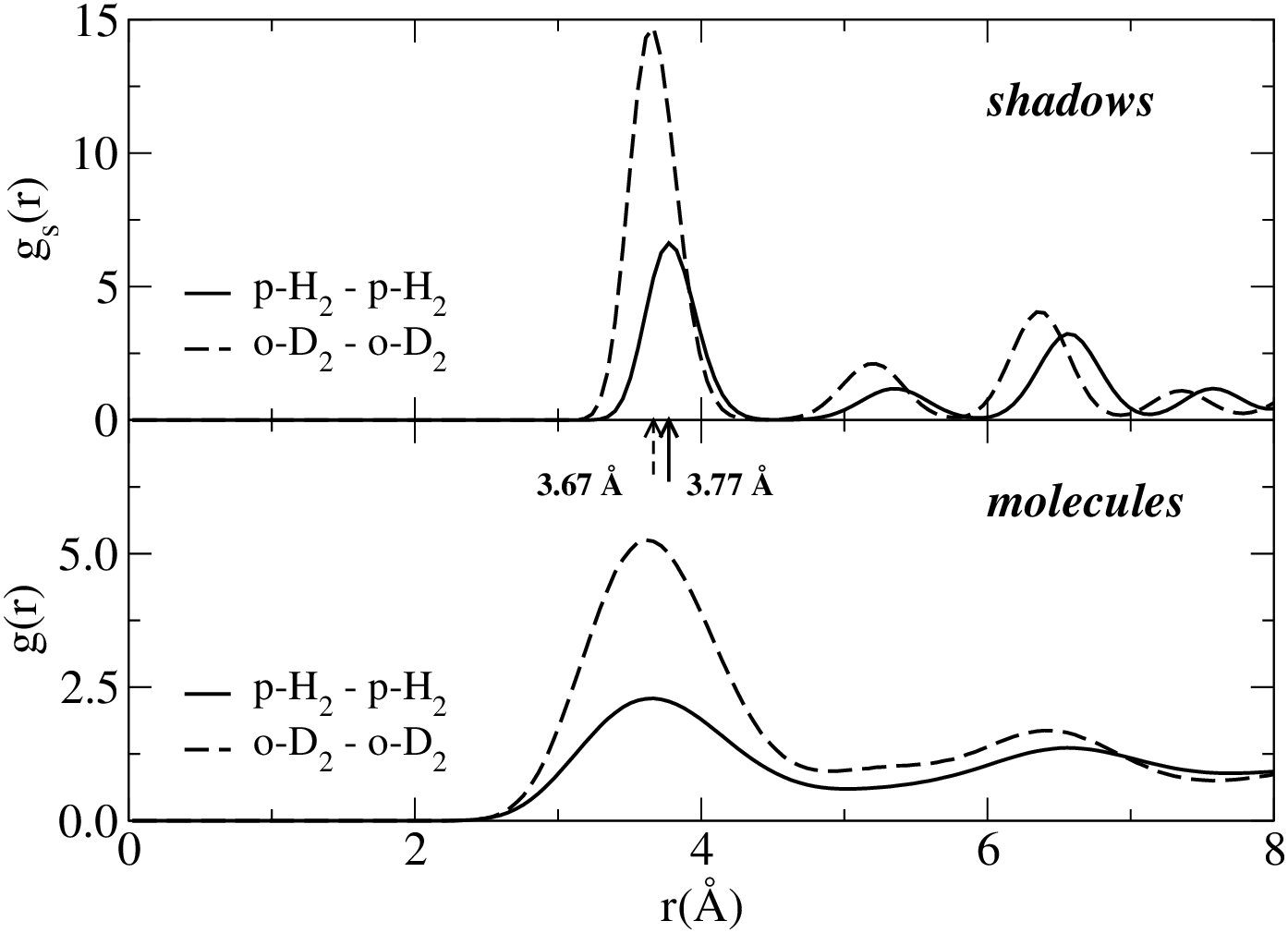}
\caption{Pair correlations functions $g_{\alpha\beta}(r)$ 
and $g_{s\alpha\beta}(r)$ with  $\alpha=\beta=$\pH\ and $\alpha=\beta=$HD  for a 25\% concentration
of clustered impurities in a fcc lattice. Solid and dashed arrows indicate the positions of the next--neighbours peaks
for \pH\ and HD molecules, respectively.} 
\label{fig:gr_hd}
\end{figure}

The energy differences between the system with clustered or diluted 
impurities, can be splitted into the mean kinetic and potential energy contributions
for \pH\ molecules and impurities, as reported in Tables \ref{tab:epot_d2} and \ref{tab:epot_hd}. 	
\begin{table}
\begin{tabular}{c|c|c|c|c}
\toprule
      & \pH\ {\scriptsize close} & \pH\ {\scriptsize far} & \oD\ {\scriptsize close} & \oD\ {\scriptsize far} \\
\hline
$V$   & -158,15(2) & -160,67(2) & -171.49(3) & -164.71(3) \\
$E_{kin}$   &   69,76(3) &   71.15(3) &   47.82(4) &   45.57(3) \\
$V+E_{kin}$ &  -88,39(2) &  -89,52(3) & -123.67(3) & -119.14(3) \\
\botrule
\end{tabular}
\caption{Potential, kinetic and total energy (in Kelvin) per particle of \pH\ and \oD\ molecules for a \pH\ crystal with
25\% of \oD\ impurities. Labels ``far'' and ``close''indicate systems with diluted or clustered impurities respectively.}
\label{tab:epot_d2}
\end{table}
\begin{table}
\begin{tabular}{c|c|c|c|c}
\toprule
      & \pH\ {\scriptsize close} & \pH\ {\scriptsize far} & HD {\scriptsize close} & HD {\scriptsize far} \\
\hline
$V$   & -158,05(2) & -159,63(2) & -167.03(3) & -161.78(3) \\
$E_{kin}$   &   70,01(3) &   70.73(3) &   57.51(4) &   55.16(4) \\
$V+E_{kin}$ &  -88,04(2) &  -88,90(2) & -109.52(3) & -106.62(3) \\
\botrule
\end{tabular}
\caption{Potential, kinetic and total energy (in Kelvin) per particle of \pH\ and HD molecules for a \pH\ crystal with
25\% of HD impurities. Labels ``far'' and ``close''indicate systems with diluted or clustered impurities respectively.}
\label{tab:epot_hd}
\end{table}
From this analysis it is possible to see how  the lower energy of the system with
clustered impurities is due to the drastic lowering of the expectation of the  potential 
energy of the impurities themselves.
As shown below, this result is related to the stronger localization 
of the impurities around equilibrium positions on a lattice with
spacing reduced compared to that of  \pH\ molecules. 
Although the average total energy of the \pH\ molecules becomes
higher when the impurities are clustered, the net effect is  lower 
total energy of the system.
Some information about the distortions in the crystal induced by the 
impurities in the \pH\ matrix can be obtained by computing the pair 
correlation function
\begin{equation}
g_{\alpha\beta}(r)=\left\langle\sum_{i\neq j}\delta
\left(\vert\mathbf{r}_i^\alpha-\mathbf{r}_j^\beta\vert-r\right)\right\rangle,
\end{equation}
where $\alpha$ and $\beta$ indicate the species among which distances are computed
 and  from the configurations of the system sampled in the MC simulations.
When using the SWF it is also possible to define the same operator for the shadow degree of freedom
\begin{equation}
g_{s\alpha\beta}(s)=\left\langle\sum_{i\neq j}\delta\left(\vert\mathbf{s}_i^\alpha-\mathbf{s}_j^\beta\vert-s\right)\right\rangle.
\end{equation}
As discussed elsewhere\cite{Vit88,Ope04}, the shadows in quantum crystals
are more rigid degrees of freedom, 
which act as heavier particles indicating a sort of ``average"  position for the real quantum particles.
%The g$_{\rm {pH}_2-{\rm imp}}$(r) computed for isolated impurities compared with the  
% g$_{\rm {pH}_2-{\rm pH}_2}$(r) shows that the lattice tends to slightly relax towards
%the impurities. The shift of the first neighbors peak is of XX\AA\  and YY\AA\ respectively
%for \oD\ and HD. 
In Figures \ref{fig:gr_d2} and \ref{fig:gr_hd} we report the pair correlations functions both for \pH\
molecules and for the impurities in the case of a concentration of 25\% of clustered impurities
embedded in a fcc \pH\ matrix.
The \oD\ impurities tend to be more localized than the \pH\ molecules, as expected due the higher mass. Moreover, 
the peaks of the pair correlation function of clustered \oD\ impurities are clearly shifted inwards, with respect to 
the pair correlation of surrounding hydrogen molecules. The strong localization of the impurities around equilibrium 
positions closer to each other than ones of \pH\ give probably rise to the drastic lowering of the potential energy of 
the \oD\ molecules discussed above, making the clustering of impurities energetically favourite.


%\section{Conclusions}
In conclusion, we used a Shadow Wave Function in connection with Variational Monte Carlo calculations to study
the properties of \oD\ and HD molecules in a \pH\ crystal. In particular we computed the excess energy 
for clustered and isolated impurities at different concentrations, trying to determine a limit for the 
solubility of \oD\ and HD in \pH.

For concentrations up to 10\% and 15\%, for \oD\ and HD molecules respectively, the excess energy 
computed for clustered impurities becomes lower than that for diluted ones.

Calculations of the pair correlation function show that at large concentrations clustered 
impurities are localized around equilibrium positions closer than ones of \pH\ molecule
giving rise to a drastic lowering of the potential energy of the impurities  themselves.

%\begin{acknowledgments}
We are grateful to C.\,J.\,Umrigar and P.\,Nightingale for providing us 
the Levemberg-Marquardt minimization code used for optimizing the parameters 
in the variational wave function.

%Calculations for this work have been partly performed at the computing facilities at ECT* in Trento.
Calculations have been performed partly on the HPC facility 
of the Departement of Physics, University of Trento, and partly at the computing facilities at ECT*
in Trento.
%\end{acknowledgments}

\end{document}